\documentclass{article}
\usepackage[utf8]{inputenc}
\usepackage{times}
\usepackage{amsfonts}
\usepackage{amsmath,amssymb}
\usepackage{indentfirst}
\usepackage{wrapfig}
\usepackage{graphicx}
\usepackage{graphics}
\usepackage{caption}
\usepackage{subcaption}
\usepackage[toc,page]{appendix}
\usepackage{color}
\usepackage{cite}
\usepackage[inactive]{srcltx}
\usepackage[T1]{fontenc}
\usepackage{float}
\usepackage{hyperref}

\begin{document}
	\date{}
\begin{center}
	{\Large\bf Enhancing nonclassical properties of quantum states of light using linear optics}
\end{center}
\begin{center}
	{\normalsize E.P. Mattos and A. Vidiella-Barranco \footnote{vidiella@ifi.unicamp.br}}
\end{center}
\begin{center}
	{\normalsize{Gleb Wataghin Institute of Physics - University of Campinas}}\\
	{\normalsize{ 13083-859   Campinas,  SP,  Brazil}}\\
\end{center}
\begin{abstract}
In this letter, we present a simple and versatile scheme for enhancing the nonclassical properties of light states using 
only linear optics and photodetectors. By combining a coherent state $|\alpha\rangle$ and an arbitrary pure state of 
light $|\phi\rangle$ (excluding coherent states) at two beam splitters, we show that the amplitude $\alpha$ of the coherent 
state can be tuned to filter out specific Fock components and generate states of light with enhanced nonclassical features. 
We provide two examples of input states and demonstrate the effectiveness of our scheme in enhancing the sub-Poissonian 
statistics or the quadrature squeezing of the output states.
\end{abstract}
%
\section{Introduction}
In recent years, the field of quantum technologies \cite{barnett17} has made significant strides, yet there are still challenges 
that must be overcome for these technologies to be effectively deployed. An essential requirement for leveraging such technologies is the 
preparation of states of light having nonclassical properties like quadrature squeezing or sub-Poissonian statistics. Several quantum 
state engineering protocols can be found in the literature, involving the interaction of light in special media \cite{kilin95,avb98a}, 
or in cavity QED systems \cite{schleich93,eberly96}, as well as schemes of photon addition \cite{agarwal91,panigrahi21,honarasa23}, 
photon subtraction \cite{agarwal92}, the use of arrays of beam splitters followed by photodetections \cite{welsch99}, quantum state 
truncation ("quantum scissors") \cite{barnett98,avb21,avb22}, and Fock state filtering ("hole burning") \cite{baseia98,baseia04,zeilinger06}. 
Those methods can provide a large variety of quantum states of light but often require nonclassical resources, e.g., 
the previous generation of single-photon states \cite{barnett98}. Depending on the intended application, it is important to enhance 
specific nonclassical features of a given state. Therefore, there is great interest in exploring new methods for generating nonclassical 
states of light, particularly those that offer more simplified setups. 

\begin{figure}[h]
	\centering
	\includegraphics[scale=0.50]{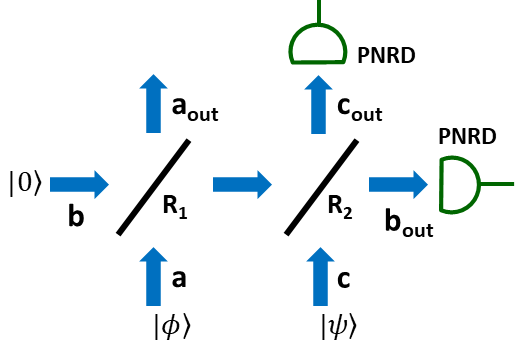}
	\caption{Schematic illustration of the proposed setup: two beam splitters (reflectances $\mathtt{R}_1$ and $\mathtt{R}_2$)
		are placed sideways. A state $|\phi\rangle$ is injected in port $a$, the vacuum state $|0\rangle$ in port $b$, and in port 
		$c$ it is injected a state $|\psi\rangle$.}
	\label{fig:setup_hole_burning}
\end{figure}

In this letter, we propose an arrangement that includes two beam splitters and two photodetectors, 
similar to the traditional quantum scissors setup \cite{barnett98}, but that will be employed for a completely different purpose. In place of having a (highly nonclassical) 
single-photon state $|1\rangle$ as input into one of the ports in the first beam splitter, we consider that a pure state $|\phi\rangle$, other than a coherent state 
enters such a port, i.e., a state that may have 
some degree of nonclassicality \cite{dodonov02}. Meanwhile, a (quasi-classical) coherent state $|\alpha\rangle$ is simultaneously injected into the input port of a 
second beam splitter, and two of the emerging beams impinge upon photodetectors. This configuration does not lead to state truncation, as happens in 
the original scissors, but rather, quantum interference allows us to perform Fock state filtering. By selecting an appropriate amplitude $\alpha$ for 
the auxiliary coherent state, it becomes possible to generate states with specific Fock components removed. The fact that our method uses the assistance of  
readily accessible coherent states represents a clear advantage over other methods of "hole burning" presented in the literature, which either require 
additional nonlinear interactions (e.g., a Kerr medium \cite{gerry02}), a cavity QED setup \cite{yang18}, or the previous generation of single-photon 
ancilla states \cite{baseia04}. In contrast, our scheme demands only linear optics and photodetectors, enabling straightforward tuning via the auxiliary 
coherent states. Furthermore, we will show that the resulting filtered states may have some of their 
non-classical features enhanced compared to the properties of the original input states $|\phi\rangle$. 
The arrangement we are here proposing for Fock state filtering requires two beam splitters ($j = 1,2$) with transmittances $\mathtt{T}_j = \cos\theta_j$ 
and reflectances $\mathtt{R}_j = i\sin\theta_j$, and two photodetectors, as shown in Fig.\ref{fig:setup_hole_burning}. We assume that 
an arbitrary state $|\phi\rangle = \sum_{i=0}^\infty \phi_i |i\rangle$ is injected in port $a$, in place of a single-photon state as it is done in the 
traditional quantum scissors. In port $b$ we have the vacuum state $|0\rangle$, and in port $c$ it is injected another state 
$|\psi\rangle = \sum_{n=0}^\infty \psi_n |n\rangle$. As already mentioned, the state $|\phi\rangle$ can be any quantum pure state of light 
excluding the coherent states. This is because when two light beams prepared in coherent states are mixed in a beam splitter, they do not become 
entangled, which invalidates the scheme. The overall input state is then given by 
$|in\rangle = |\phi\rangle|0\rangle|\psi\rangle = \sum_{i=0}^\infty\sum_{n=0}^\infty\phi_i\psi_n|i, 0, n\rangle$, and the resulting output 
state after crossing the beam splitter arrangement will be
\begin{eqnarray}
	|out\rangle = \hat{R}_2\hat{R}_1|in\rangle=
	\sum_{i=0}^\infty\sum_{n=0}^\infty\frac{1}{\sqrt{i!}}\frac{1}{\sqrt{n!}}\phi_i\psi_n\hat{R}_2\hat{R}_1\hat{a}^{\dag i}\hat{c}^{\dag n}|0, 0, 0\rangle \\ \nonumber
	= \sum_{i=0}^\infty\sum_{n=0}^\infty\sum_{j=0}^i\sum_{k=0}^j\sum_{m=0}^n\frac{\sqrt{i}}{j!(i-j)!}\frac{\sqrt{n}}{m!(n-m)!}\frac{j!}{k!(j-k)!} \\ \nonumber
	\times \sqrt{(i-j)!}\sqrt{(j-k+n-m)!}\sqrt{(k+m)!} \phi_i\psi_n \\ \nonumber
	\times \mathtt{T}_1^{i-j}(-\mathtt{R}_1^*)^j \mathtt{T}_2^{*j-k} \mathtt{R}_2^k(-\mathtt{R}_2^*)^{n-m} \mathtt{T}_2^m|i-j, j-k+n-m, k+m\rangle,
\end{eqnarray}
being $\hat{R}_1 = \exp\left[i\theta_1(\hat{a}^\dagger\hat{b} + \hat{a}\hat{b}^\dagger)\right]$ and  
$\hat{R}_2 = \exp\left[i\theta_2(\hat{c}^\dagger\hat{b} + \hat{c}\hat{b}^\dagger)\right]$ the beam splitter operators.

If photodetectors are now placed in output ports $b$ and $c$, and $1$ and $0$ photons are detected, respectively, the output state arising
in port $a$ will be
\begin{equation}
	|H\rangle = \frac{1}{\sqrt{p}}\sum_{i=0}^\infty \mathtt{T}_1^i(\phi_i\psi_1 \mathtt{R}_2^*+\sqrt{i+1}\phi_{i+1}\psi_0\mathtt{R}_1^* \mathtt{T}_2^*)|i\rangle,
	\label{collapsedoutputstate}
\end{equation}
with a probability of generation
\begin{equation}
	p=\sum_{i=0}^\infty |\mathtt{T}_1^i|^2|\phi_i\psi_1 \mathtt{R}_2^*+\sqrt{i+1}\phi_{i+1}\psi_0 \mathtt{R}_1^*\mathtt{T}_2^*|^2.
\end{equation}
As a matter of fact, the whole process can be viewed as the action of an operator $\hat{\cal H}$ on the input state 
$|\phi\rangle$, or $|H\rangle \propto \hat{\cal H} \sum_i \phi_i|i\rangle$, being
\begin{equation}
	\hat{\cal H} = T_1^{\hat{a}^\dagger \hat{a}}\left(\hat{a} + \frac{\psi_1}{\psi_0 \Lambda}\hat{I}\right),
\end{equation}
and where $\Lambda = \mathtt{R}_1^* \mathtt{T}_2^*/\mathtt{R}_2^*$.
This corresponds to a coherent superposition of a one-photon subtraction with a zero-photon subtraction process, with the latter sometimes 
referred to as "noiseless attenuation" in the literature \cite{cerf12,franson22}. Likewise, there have been proposed schemes of quantum state
engineering, to generate arbitrary finite superpositions of Fock states via repeated photon subtractions and squeezing \cite{cerf05}, as well as 
to produce continuous-variable qubit states \cite{sasaki10}, a type of Schr\"odinger cat states \cite{dodonov74,avb92}.
Now, from Eq.(\ref{collapsedoutputstate}) we note that the vacuum component $|0\rangle$ can be removed from the state $|H\rangle$ if the relation 
\begin{equation}
	\phi_0\psi_1 \mathtt{R}_2^*+\phi_{1}\psi_0 \mathtt{R}_1^* \mathtt{T}_2^*=0
	\label{condition}
\end{equation}
is satisfied.
In particular, if the state $|\psi\rangle$ is a coherent state, i.e., $|\psi\rangle = |\alpha\rangle$, the condition in 
Eq.(\ref{condition}) will hold provided we have
\begin{equation}
	\alpha^{(0)}=- \Lambda\frac{\phi_{1}}{\phi_0}.
\end{equation}
In general, the removal of the $n$-th Fock component is accomplished as long as we have
\begin{equation}
	\alpha^{(n)}=-\Lambda\sqrt{n+1} \frac{\phi_{n+1}}{\phi_n}.\label{removalcondition}
\end{equation}
We should remark that provided the coefficients obey the relation in Eq.(\ref{removalcondition}), it becomes 
possible to remove Fock components other than the $n$-th. Hereafter, whenever we mention the removal of the $n$-th Fock component, 
this will mean that the amplitude $\alpha^{(n)}$ has been chosen in order to exactly remove such a component. Also, the "hole burning" 
as proposed here is not universal in the sense that it works only for specific states satisfying Eq.(\ref{removalcondition}). 
While previous schemes described in the literature \cite{baseia04} allow for the controlled removal of Fock components using 
beam splitters and photodetections, they typically require the prior generation of Fock states, which is not necessary in 
our proposed approach.

\begin{figure}[h]
	\centering
	\includegraphics[scale=0.45]{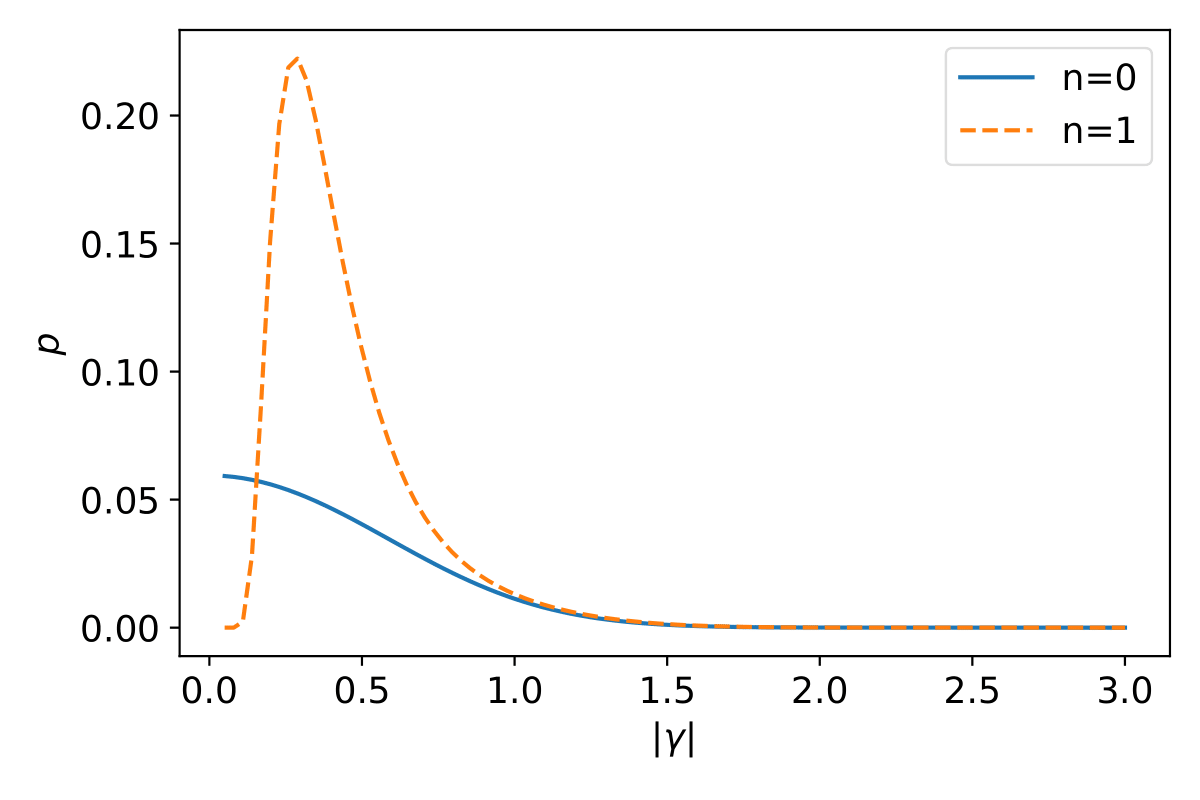}
	\caption{Generation probability of state $|h;\gamma,\xi\rangle$ as a function of $|\gamma|$,
		with $50:50$ beam splitters, $\beta = \varphi = 0$, and $s = 1.0$.}
	\label{fig:p_sqst}
\end{figure}

In the following discussion, we will examine examples of output states that can be generated from two distinct 
initial states (which are not coherent states for the reason stated above), and assume for simplicity that both beam splitters have 
a $50:50$ splitting ratio. Firstly we assume that the input state $|\phi\rangle$ 
in port $a$ is a squeezed coherent state, or
\begin{equation}
	|\gamma, \xi\rangle=\hat{D}(\gamma)\hat{S}(\xi)|0\rangle,
\end{equation}
where $\hat{D}(\gamma) = \exp(\gamma\hat{a}^\dagger - \gamma^*\hat{a})$ is the displacement operator (coherent amplitude $\gamma=|\gamma|e^{i\beta}$)
and $\hat{S}(\xi) = \exp[(\xi^*\hat{a}^2 - \xi\hat{a}^\dagger{}^2)/2]$ the squeezing operator (squeezing parameter $\xi=se^{i\varphi}$).
While the quasi-classical coherent state was brought to the optical context in the 1960s \cite{glauber63}, the 
nonclassical squeezed states were "rediscovered" in the following decade \cite{stoler70}. A prominent nonclassical
property of such states is quadrature squeezing, the reduction of noise in one of the quadrature variables,
which are defined as $\hat{X} = \left(\hat{a}+\hat{a}^{\dagger}\right)/2$ and $\hat{Y} = \left(\hat{a}-\hat{a}^{\dagger}\right)/2i$. 
The quadrature operators obey $\left[\hat{X},\hat{Y}\right]=i/2$, and consequently 
$\langle(\Delta\hat{X})^{2}\rangle \langle\Delta\hat{Y})^{2}\rangle \geq1/16$. Squeezing of quantum noise, e.g.,
in the $\hat{X}$ quadrature is verified if $\langle(\Delta\hat{X})^{2}\rangle < \frac{1}{4}$ (analogously for
the $\hat{Y}$ quadrature). The resulting output states after the photodetections, denoted here as $|h;\gamma,\xi\rangle$,  
have generation probabilities $p$ that depend on the parameters involved. In Fig.\ref{fig:p_sqst} we plotted the generation 
probability as a function of the coherent magnitude $|\gamma|$, in the cases of having completely removed either the $n = 0$ or 
the $n = 1$ component of the original state. To simplify matters, we have chosen both the phase of the coherent 
amplitude and the squeezing parameter phase as being null ($\beta = \varphi = 0.0$). We also used a squeezing parameter, 
$s = 1.0$, of sufficient magnitude to give rise to interesting effects for the generated states. 

\begin{figure}[h]
	\centering
	\includegraphics[scale=0.45]{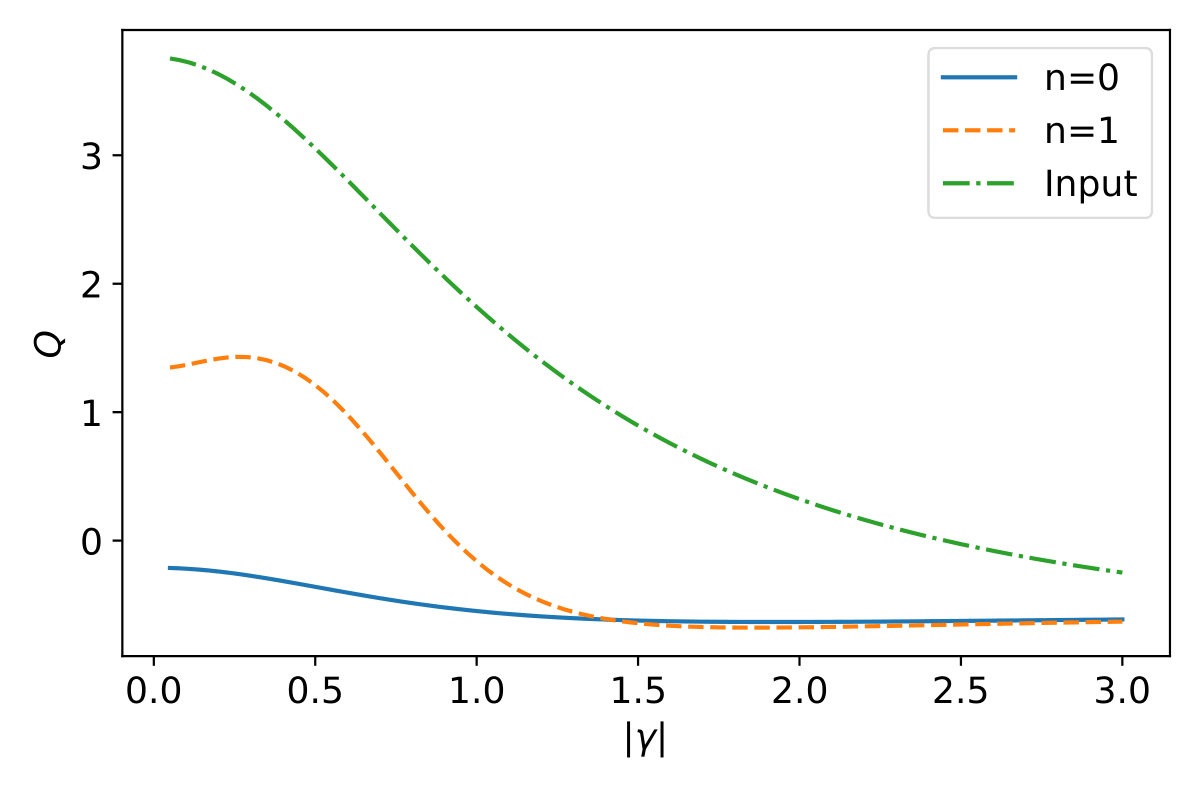}
	\caption{Mandel's $Q$ parameter of state $|h;\gamma,\xi\rangle$ as a function of $|\gamma|$,
		with $50:50$ beam splitters, $\beta = \varphi = 0$ and $s = 1.0$. The green, dot-dashed curve refers 
		to the input state $|\gamma,\xi\rangle$.}
	\label{fig:q_sqst}
\end{figure}

The states $|h;\gamma,\xi\rangle$ can have reduced photon number fluctuations compared to the
input states $|\gamma,\xi\rangle$, which in this case are super-Poissonian. This can be quantified 
via Mandel's $Q$ parameter, defined as $Q =  \left\langle (\Delta \hat{n})^2\right\rangle/\left\langle \hat{n} \right\rangle - 1$;
it has a minimum value of $Q = -1$ for Fock states and is null for coherent states. In Fig.\ref{fig:q_sqst} we have plotted the $Q$ 
parameter relative to state $|h;\gamma,\xi\rangle$ as a function of $|\gamma|$. Undoubtedly, the generated state has 
fluctuations in the number of photons reduced compared to the input state. We note that a more sub-Poissonian field is generated if 
the Fock component $n = 0$ is removed instead of the $n = 1$ one, especially in a low-intensity regime (small $|\gamma|$). This is
because if the $n = 0$ component is removed, the presence of the one-photon component enhances the sub-Poissonian character of the 
generated field. Conversely, if we remove the $n = 1$ component while the vacuum component remains, such an enhancement will be much 
less pronounced, as clearly shown in Fig.\ref{fig:q_sqst}.

\begin{figure}[h]
	\centering
	\includegraphics[scale=0.45]{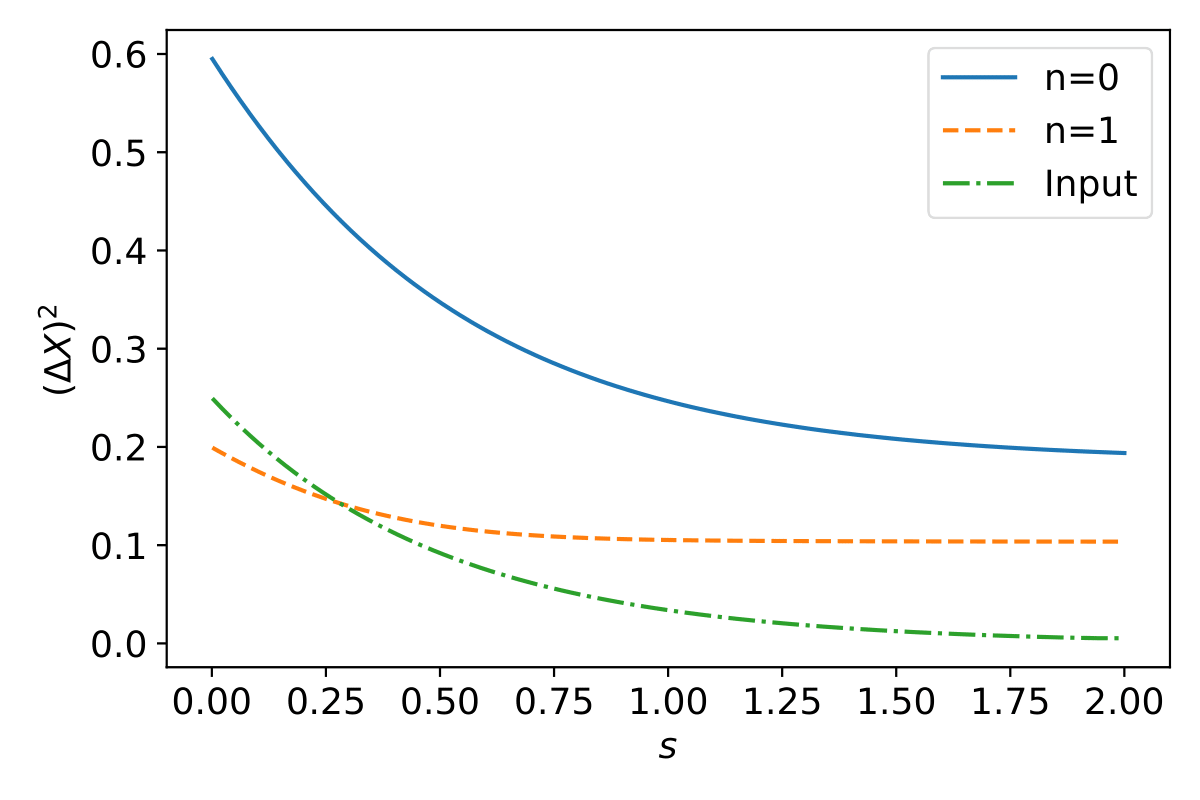}
	\caption{Variance of the $\hat{X}$ quadrature of state $|h;\gamma,\xi\rangle$ as a function of $s$,
		with $50:50$ beam splitters, $\beta = \varphi = 0$ and $|\gamma| = 0.5$. The green, dot-dashed curve refers to the 
		input state $|\gamma,\xi\rangle$.}
	\label{fig:vx_sqst}
\end{figure}

The input state $|\gamma,\xi\rangle$ being itself an ideal squeezed state, naturally exhibits quadrature squeezing. 
Nevertheless, for small values of $s$, it is possible
to have an output state $|h;\gamma,\xi\rangle$ slightly more squeezed than the input state if the 
$n = 1$ Fock component is removed. Yet, this occurs only for smaller values of the 
generation probability $p$, as shown in Fig.\ref{fig:vx_sqst}, where we have
plotted the variance of the $\hat{X}$ quadrature as a function of $s$.

\begin{figure}[h]
	\centering
	\includegraphics[scale=0.45]{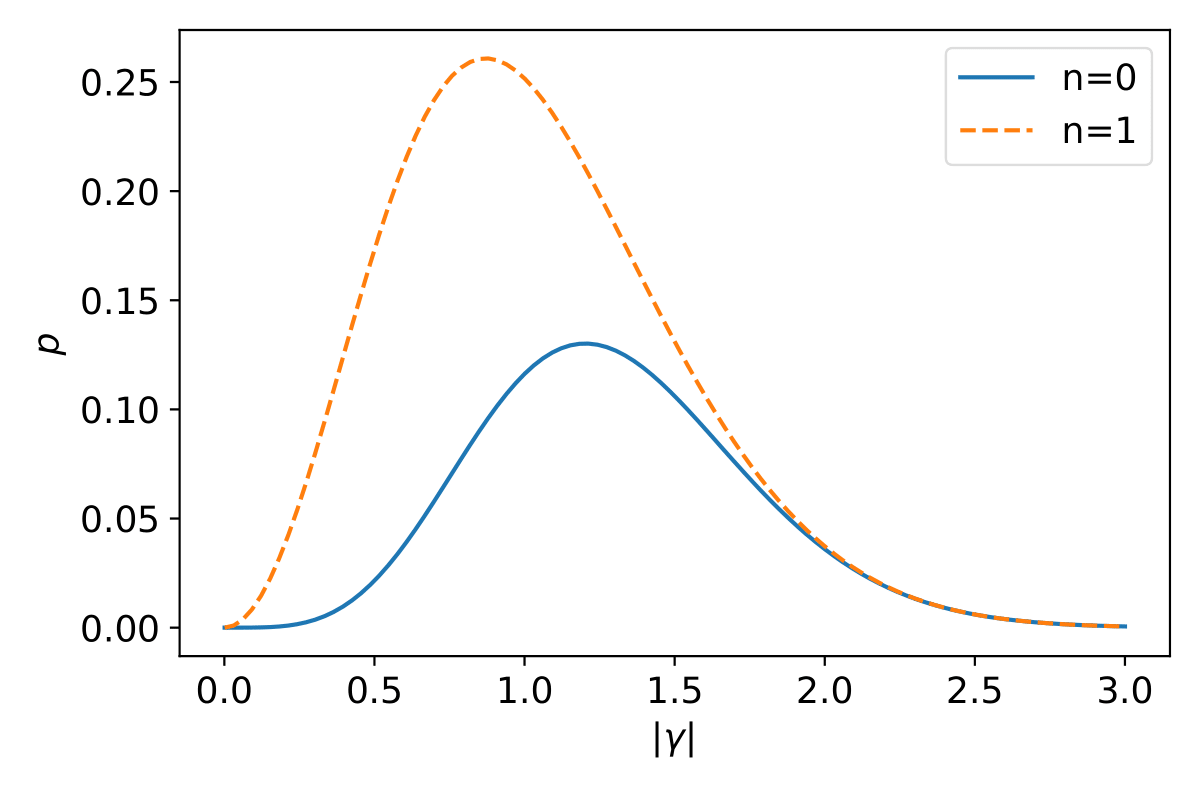}
	\caption{Generation probabilities of odd states $|h;\gamma,-\rangle$ (blue, continuous curve) and
		even states $|h;\gamma,+\rangle$ (orange, dashed curve) as a function of $|\gamma|$ with 
		$50:50$ beam splitters, $\beta = 0$ and $\delta = \pi/2$.}
	\label{fig:p_cat}
\end{figure}

\begin{figure}[h]
	\centering
	\includegraphics[scale=0.45]{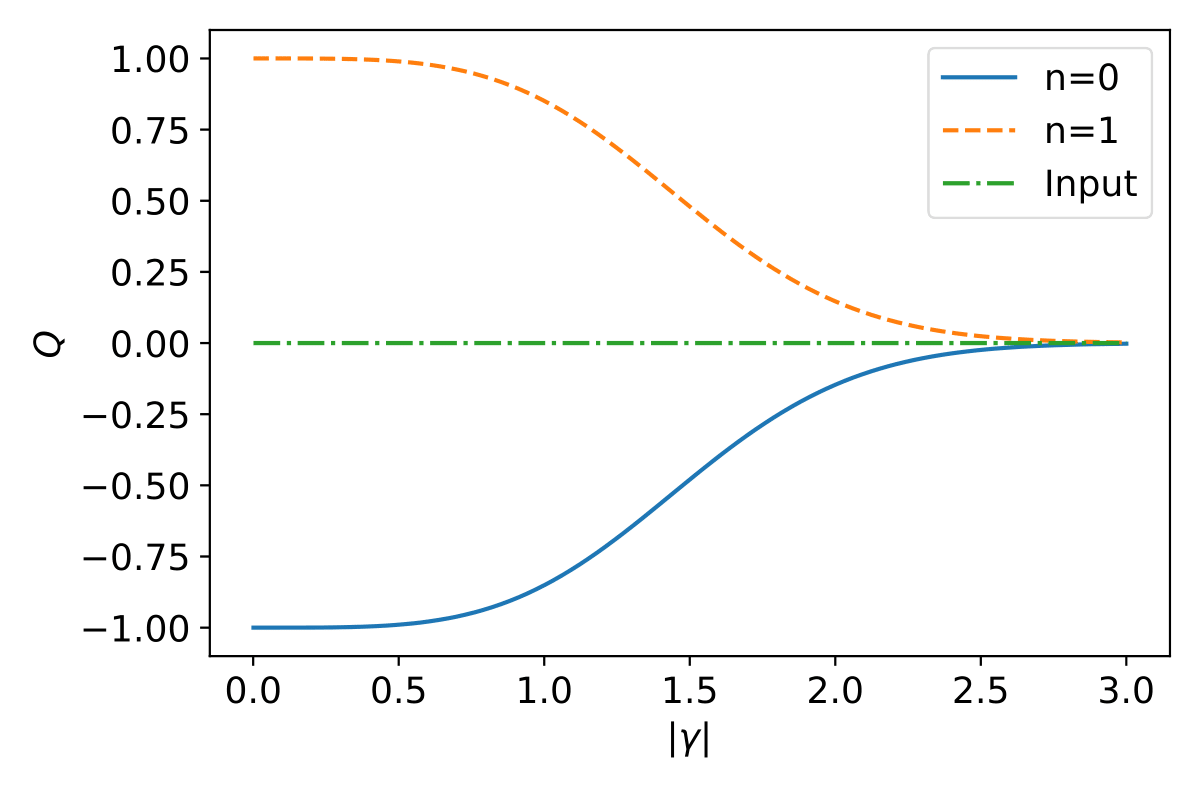}
	\caption{Mandel's $Q$ parameter of odd states $|h;\gamma,-\rangle$ (blue, continuous curve) and
		even states $|h;\gamma,+\rangle$ (orange, dashed curve) as a function of $|\gamma|$ with 
		$50:50$ beam splitters, $\beta = 0$ and $\delta = \pi/2$. The green, dot-dashed curve refers to the input state, 
		$|\gamma,\delta\rangle$.}
	\label{fig:q_cat}
\end{figure}

\begin{figure}[h]
	\centering
	\includegraphics[scale=0.45]{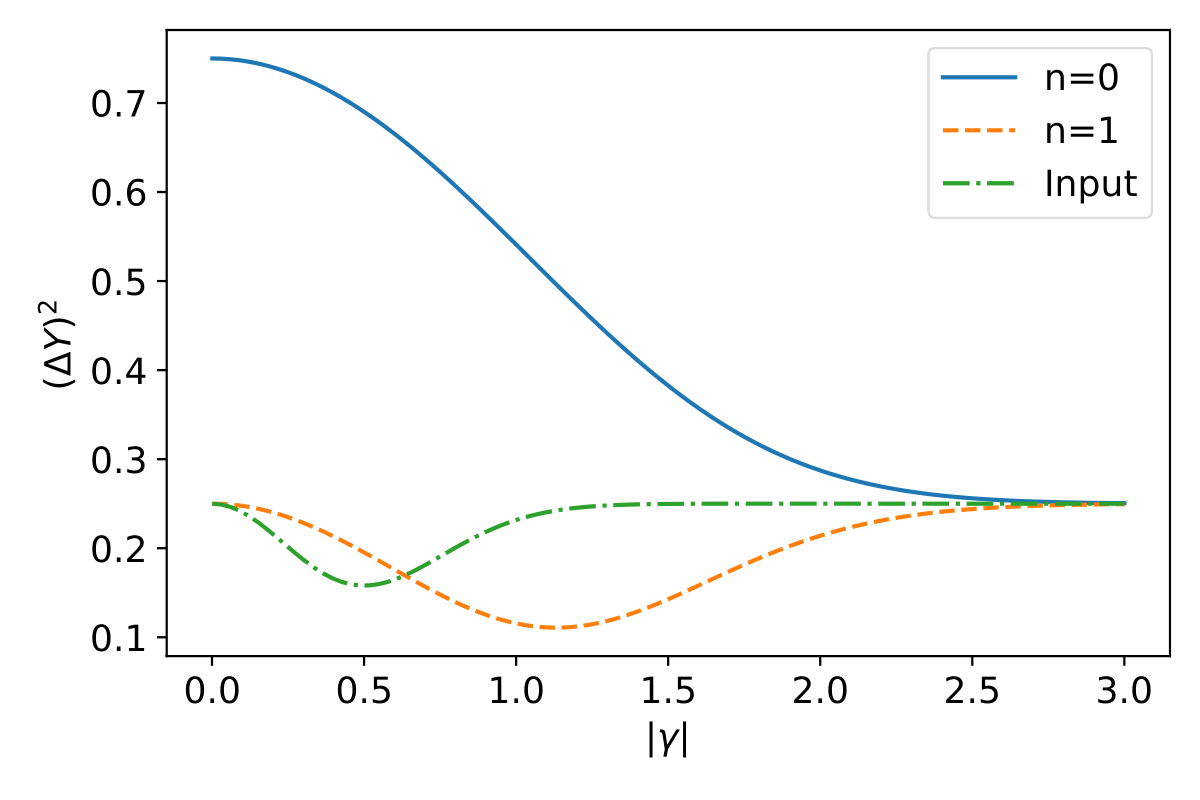}
	\caption{Variance of the $\hat{Y}$ quadrature of odd states $|h;\gamma,-\rangle$ (blue, continuous curve) and
		even states $|h;\gamma,+\rangle$ (orange, dashed curve) as a function of $|\gamma|$ with $50:50$ beam splitters, 
		$\beta = 0$ and $\delta = \pi/2$. The green, dot-dashed curve refers to the input state, $|\gamma,\delta\rangle$.}
	\label{fig:vy_cat}
\end{figure}

Another state of light presenting nonclassical properties is a quantum superposition of two coherent states, $|\gamma\rangle$ and 
$|-\gamma\rangle$, also known as a Schr\"odinger cat state \cite{dodonov74,avb92}
\begin{equation}
	|\gamma, \delta\rangle = C (|\gamma\rangle + e^{i\delta}|-\gamma\rangle),\label{catstate}
\end{equation}
where $\delta$ is a relative phase and $C$ the normalizing constant. The cat states can exhibit a variety of nonclassical features
depending on the phase $\delta$ \cite{avb92}. For instance, they can be sub-Poissonian ($\delta = \pi$), Poissonian ($\delta = \pi/2$)
or super-Poissonian ($\delta = 0$). We have that in this case, there are two possible scenarios for the resulting states $|H\rangle$: either all even 
or all odd Fock components will be removed. In other words, to generate a state having only even Fock components, $|h;\gamma,+\rangle$, 
the amplitude of the auxiliary coherent state $|\psi\rangle$ has to be
\begin{equation}
	\alpha^{(even)} = -\gamma \Lambda \frac{1+e^{i\delta}}{1-e^{i\delta}},
\end{equation}
while for an output state having only odd Fock components, $|h;\gamma,-\rangle$, the coherent amplitude should be
\begin{equation}
	\alpha^{(odd)} = -\gamma \Lambda \frac{1-e^{i\delta}}{1+e^{i\delta}}.
\end{equation}
In this context, a convenient choice for the relative phase in the cat state is $\delta = \pi/2$ (Yurke-Stoler state). In Fig.\ref{fig:p_cat}
we have plots of the probability generation of the states $|H\rangle$ as a function of $|\gamma|$. It is worth mentioning that 
in the following plots, the notation $n = 0$ ($n = 1$) indicates the removal of all even (odd) Fock components. A remarkable effect occurs in 
relation to the photon number statistics of the generated states. The input Yurke-Stoler state has Poissonian statistics, but depending on the 
parity chosen, the resulting states will have very different properties. As shown in Fig.\ref{fig:q_cat}, the states $|h;\gamma,+\rangle$ are basically 
super-Poissonian, while the states$|h;\gamma,-\rangle$ are sub-Poissonian. Note that for large enough $|\gamma|$, both states become
Poissonian ($Q\rightarrow 0$). Differently from the other two types of states discussed above, the cat state in Eq.(\ref{catstate}) may present 
squeezing of the $\hat{Y}$ quadrature, instead. This is shown in Fig.\ref{fig:vy_cat}, where we have plotted $\langle(\Delta\hat{Y})^{2}\rangle$ as
a function of $|\gamma|$ for the cat state together with states $|h;\gamma,\pm\rangle$. Interestingly, while the generated states $|h;\gamma,-\rangle$
do not present squeezing, the states $|h;\gamma,+\rangle$ may have a higher degree of squeezing than the original cat states.

In conclusion, we proposed a method of tunable Fock state filtering based on linear optics and photodetections. We demonstrate that the quantum interference between a 
quasi-classical coherent state $|\alpha\rangle$ and a state $|\phi\rangle$ (other than a coherent state) in an arrangement containing two beam splitters and followed by 
photodetections, enables the removal of specific Fock components from previously prepared states. Distinct nonclassical states can be generated simply by adjusting the 
amplitude $\alpha$ of the ancilla coherent state. In order to showcase the effectiveness of our method, we considered two different input states: a squeezed 
coherent state and a Schr\"odinger cat state. Our findings reveal that Fock-filtered states derived from these input states can exhibit 
enhanced nonclassical properties compared to the original states.


\bibliographystyle{unsrt}
\bibliography{refsfilter}


\section*{Acknowledgments}
 E.P.M. would like to acknowledge financial support from CAPES (Coordenadoria de Aperfeiçoamento
 de Pessoal de Nível Superior), Brazil, grant N${\textsuperscript{\underline{o}}}$ 88887.514500/2020-00.
 This work was also supported by CNPq (Conselho Nacional para o
 Desenvolvimento Científico e Tecnológico, Brazil), via the INCT-IQ
 (National Institute for Science and Technology of Quantum Information),
 grant N${\textsuperscript{\underline{o}}}$ 465469/2014-0.



\end{document}